# Title: One photon-per-bit receiver using near-noiseless phase-sensitive amplification

**Authors**: Ravikiran Kakarla, Jochen Schröder and Peter A. Andrekson*

*****Corresponding author. Email**: peter.andrekson@chalmers.se

**Affiliations**: Photonics laboratory, Department of Microtechnology and Nanoscience, Chalmers University of Technology, Gothenburg, Sweden

**One sentence summary**: Approaching the ultimate reach in space communications.

**Abstract:**

Noise fundamentally limits the capacity and reach in all communication links. In optical space communications, noise primarily originates from the detection process and limits the signal fidelity. . Therefore, the receiver sensitivity plays a key role, dictating the minimum power needed to recover the information transmitted. The widely explored approach of using the pulse-position modulation format trades-off sensitivity against receiver bandwidth and thus data-rate. Here we report on a novel, spectrally efficient, approach based on a coherent receiver with a near-noiseless phase-sensitive pre-amplifier operating at room temperature and demonstrate a sensitivity of one photon-per-bit of incident power at a data rate of 10 Gb/s. The results provide a path to future high-capacity inter-satellite and deep space, and other free-space communication links.

**Main Text:**

Increased space exploration and the growing capability and thus data output of satellite born sensors operated by agencies like NASA, ESA and JAXA imposes greater demands on the communication systems to operate at higher data rates and to reach farther distances into space (*1*). Improving the receiver sensitivity is considered the most important method to improve data throughput with as few received photons as possible. A better receiver sensitivity translates to longer reach, higher data throughput and the ability to use more compact optics or a combination of the above. Common approaches to improve the sensitivity are known to suffer from poor spectral efficiency (bits/s per Hz) and achieve only modest net data rates due to their inherent trade-off between sensitivity and bandwidth (*2*). In particular, PPM is widely considered in space communications as it can reach excellent sensitivity at very low signal-to-noise ratio (SNRs) (*3*), but with significant loss of spectral efficiency. Photon counting receivers are often used to receive PPM symbols enabling sensitivities of few photons per bit. Superconducting nanowire-based versions of these have recently been shown to provide excellent performance including a quantum efficiency of 90% at data rates up to a few 100 Mb/s (*4–6*). A key drawback is their need to be cooled to 2-4 K, making it difficult to deploy in satellites and their inability to detect photons at rates of multiple Gb/s.

The record sensitivity demonstrations of photon counting receivers with PPM modulation include a sensitivity of approximately 1 PPB at 14 Mbps (*7*), and 1.2 PPB at 38 Mbps (*8*). However, the low efficiencies of photon counting receivers at high frequencies results in relatively modest performance at higher data rates. Successful application of PPM and photon counting technology at rates above 100 Mbps have been demonstrated by NASA at 622 Mbps with a 3.8 PPB sensitivity in the Lunar Laser Communication Demonstration (LLCD) (*8*) and at 781 Mbps with a sensitivity of 0.5 detected PPB. However, when accounting for insertion loss and non-ideal quantum efficiency this latter result corresponds to a "black box" sensitivity of approximately 8 incident PPB (*9*).

Future space communication systems such as inter-satellite links and satellite to ground are expected to operate at speeds of several 10s of Gb/s and beyond and will thus require a significant improvement over existing receiver technology both in terms of data rate and sensitivity (*6*). Space communication research has therefore started to adopt technology from optical fiber communication, and advanced modulation formats with optically pre-amplified coherent detection in combination with advanced FEC are a promising solution to improve both data rates and receiver sensitivity. One impressive result using single-quadrature homodyne detection receiver without pre-amplifier resulted in a sensitivity of 1.5 PPB at 156 Mb/s (*10*), where the data rate was limited by the optical PLL bandwidth. Demonstrations with EDFA pre-amplified coherent receivers resulted in sensitives as low as 2.1 PPB at 10 Gb/s (*11–13*).

Here we demonstrate a novel use of an ultralow noise phase-sensitive optical amplifier as pre-amplifier in a coherent receiver resulting in an unprecedented "black-box" sensitivity of 1 PPB with error-free performance at 10.5 Gb/s. This is not only much more spectrally efficient than the PPM approaches allowing the use of very high data rates, but also operates at room temperature.

The theoretically minimal sensitivities are found from inspecting the fundamental capacity limits. The capacity, i.e. the maximum information rate with error-free data transmission for a pre-amplified dual-quadrature or phase-diverse coherent homodyne receiver is (*14*) [see the supplementary for derivation]:

$$C_{preamp} = B \log_2 \left(1 + \frac{2S}{F_N h \nu B}\right)$$

where $F_N$ is the noise figure of the pre-amplifier, $S$ is the signal power, $h$ is Planck's constant, $\nu$ is the frequency of the optical carrier wave, and the bandwidth $B$ is the inverse of the symbol period. By rewriting $S = n_s h \nu B$, where $n_s$ is the number of photons per transmitted symbol, the capacity of a pre-amplified receiver becomes $C_{preamp} = B \log_2 \left(1 + 2n_s / F_N\right)$ with $2n_s / F_N$ interpreted as the SNR of the signal. For an erbium-doped fiber amplifier (EDFA), the best possible noise figure is 3 dB, and therefore $C_{EDFA} = B \log_2 \left(1 + n_s\right)$. The same expression is reached in a shot-noise limited dual-quadrature receiver assuming 100% detector quantum efficiency (*14*).

In contrast to EDFAs, PSAs have a theoretical noise figure of 0 dB, amplifying the signal without excess noise (*15*). A PSA that can amplify both quadratures of an optical wave, requires two input waves at different wavelengths, the signal and its conjugate (idler).

Due to the coherent addition of input waves and incoherent addition of noise by four-wave mixing in the PSA, the output SNR of each wave is 3 dB better than the input, corresponding to a noise figure of -3 dB. However, the overall noise figure of the PSA is still 0 dB when accounting for both the required input waves with the same information (*16*).

Due to the four-fold SNR improvement over the EDFA (which degrades the SNR by 3 dB), the capacity of the PSA is $C_{PSA} = \frac{B}{2}\log_2(1+4n_s)$ where the factor ½ is due to the loss of spectral efficiency as signal and idler are carrying the same information [see the supplementary for detailed derivation].

In the limit of SNR $\rightarrow 0$, the capacity of the PSA amplified receiver is $C_{PSA} = 2Bn_s/\ln 2$, double that of the EDFA, $C_{EDFA} = Bn_s/\ln 2$. The ultimate sensitivity can be found by taking the ratio of the number of photons per symbol $n_s$ and the number of bits per symbol (C/B), which is also equal to the spectral efficiency, (bits/s per Hz) resulting in the best possible sensitivity of 0.35 PPB for the PSA and 0.7 PPB for the EDFA. In this work, we show experimentally PSAs can reach a black box sensitivity of 1 PPB, whose preliminary results were presented in our previous publication (*17*)

**Experiment**

Fig. 1 shows a conceptual diagram of a free-space optical transmission link with a PSA pre-amplified receiver. At the transmitter, binary data was FEC-encoded using a code from the digital video broadcasting standard (DVB-S2), consisting of a concatenation of an inner ½-rate low-density parity-check code (LDPC) an outer high-rate (0.6%) Bose-Chaudhuri-Hocquenghem (BCH) code. The data was modulated onto the signal with quaternary-phase-shift-keying (QPSK) modulation at a symbol rate of 10.52 Gbaud, resulting in a net information rate of 10.52 Gb/s. The signal was then combined with a continuous-wave pump in a "copier stage" to generate a conjugate idler wave, containing the same information as the signal, by four-wave mixing (FWM) in a nonlinear optical fiber. All three waves, signal, idler and pump are then amplified by a booster amplifier (not used in our experiment) to the desired output power and launched into the free-space channel; in our case a short 1m free-space link implemented in the laboratory followed by an optical attenuator to emulate the beam diffraction-induced loss in a real link. It should be noted that the NF of the copier and booster amplifier do not need to be below 3 dB, in fact they can be substantially higher without causing any sensitivity degradation in the receiver. The reason for this is that any excess noise from the transmitter will be attenuated to a level such that quantum noise (-61 dBm at 0.1 nm bandwidth) will dominate at the receiver due to the large link loss (*16*).

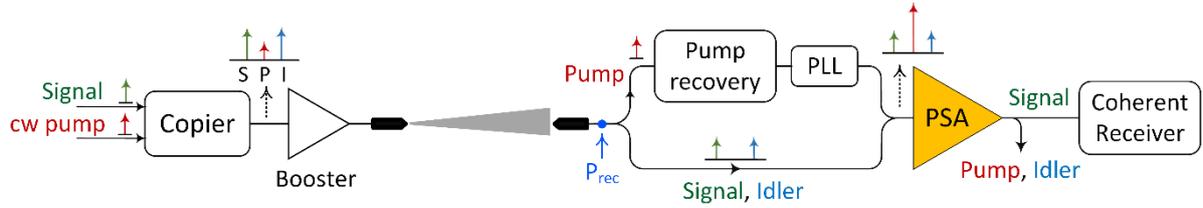

*Fig. 1. Conceptual diagram of a free-space communication link with a PSA pre-amplified coherent receiver; S-Signal; P-pump; I-Idler; PLL, Phase-locked-loop; PSA, Phase sensitive amplifier.*

The received power ($P_{rec}$), defined as the total power of the signal, idler and pump after the receiver lens as depicted in Fig. 1 thus represents the "black-box" receiver sensitivity. The launch power determined by the booster amplifier is equally important as the receiver sensitivity in an actual free-space link. It is therefore essential that the pump power corresponds to only a small fraction of the total launch power. In our case the pump power was substantially smaller than the combined signal and idler power resulting in a nearly negligible power budget penalty.

At the receiver, the pump was separated from the signal and idler with a wavelength division multiplexer (WDM) and recovered using optical injection locking (*18*). We were able to recover a stable high power (approximately 1 W) pump wave at input power levels as low as -72 dBm, which is at least 12 dB smaller than the received signal power level. The amplification of the pump was thus > 100 dB.

An optical phase-locked loop (PLL) after pump recovery maintained a constant relative phase between the three waves for maximum phase sensitive gain. The signal, idler and recovered pump were then combined inside an HNLF for phase sensitive amplification of the signal. After the PSA, the signal was filtered and detected using a standard coherent receiver and a real-time oscilloscope for subsequent off-line digital signal processing (DSP). As the idler is not needed for the detection, the receiver bandwidth requirement is same as that of an EDFA pre-amplified receiver operating at the same symbol rate.

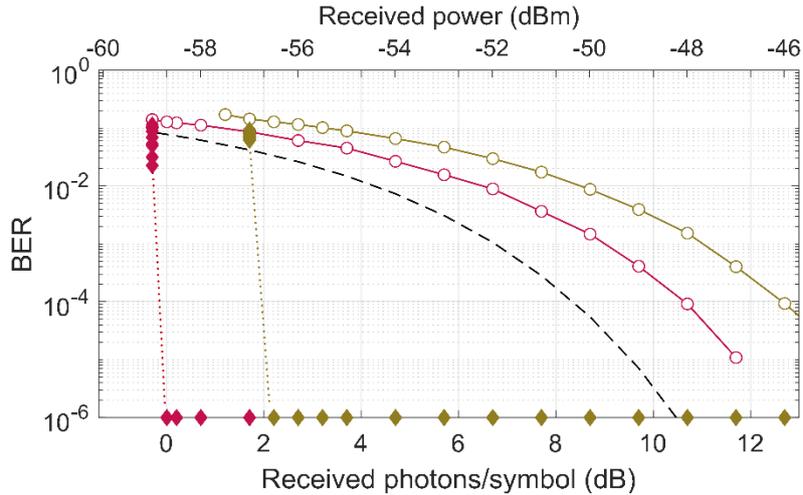

*Fig. 2. Experimental results with 10.52 Gbaud QPSK data, showing BER versus received power (also expressed in photons per symbol) before and after FEC decoding for: EDFA pre-FEC (brown line with open circles as measured points); EDFA post-FEC (brown solid diamond markers); PSA pre-FEC (red line with open circles as measured points); PSA post-FEC (red diamond markers). The dashed black line shows the theoretical estimate for QPSK data for a PSA with 0 dB noise figure. Note that the power scales represent the total receiver power, i.e. signal, idler, and pump power in the case of PSA.*

The bit error rate (BER) of the received signal was measured to evaluate the performance of our receiver and compared with an EDFA pre-amplified receiver as shown in Fig 2. The power scales include total input power, i.e. signal, idler, and pump waves for the PSA, and only signal power for the EDFA. The power budget penalty caused by the presence of the pump wave in the PSA (~12 dB below the combined power of signal and idler) was at most 0.26 dB. The pre-FEC BER in Fig. 2 show that the PSA perform 2.5 dB better than the EDFA based receiver, which is attributed to difference in noise figures of the amplifiers, measured to be 1.2 dB for the PSA and 3.7 dB for the EDFA, respectively. There was no additional power penalty due to the free-space propagation compared to a back-to-back scenario with just an attenuator.

The post-FEC BER was determined after FEC decoding and is also shown in Fig. 2. A coding gain of 11.8 dB was obtained at a BER=$10^{-5}$ for both the EDFA and PSA pre-amplified receivers. The results show that error-free (below BER = $10^{-6}$, limited by the memory of the real-time oscilloscope and offline processing) transmission can be achieved with a received power of 1 photon/symbol or 1 photon/information bit (PPB) (including FEC overhead) with a PSA pre-amplifier and is the best "black-box" sensitivity reported to date. This result is more than 3 dB better than the previously best reported sensitivity of 2.1 PPB at similar data rate and FEC (*11*). In our EDFA case, error-free performance was achieved at 1.7 PPB, 2.3 dB higher than for the PSA. The received power measurement uncertainty based on the calibrated power meter was ± 0.1 dB or equivalently ± 0.02 PPB. We estimated the possible sensitivity of our specific system with an ideal FEC using generalized mutual information (GMI) to be 0.85 PPB. A discussion on the GMI results is presented in the supplementary.

Importantly, PSA pre-amplification is compatible with other methods for further sensitivity improvements, e.g. power efficient modulation formats, spatial/spectral diversity and advanced soft-decision FEC and is furthermore straightforwardly scalable to higher bit rates as well as other wavelengths using a different nonlinear platform. It should be noted that transmitting both signal and idler results in a reduction of the spectral efficiency (SE) in the optical domain. The PSA approach therefore represents a trade-off of SE versus sensitivity somewhat similar to that of PPM modulation formats. However, in contrast to PPM, because there is no need to detect the idler wave in the PSA approach, there is no particular requirement on the electrical (or receiver-) bandwidth versus bit-rate.

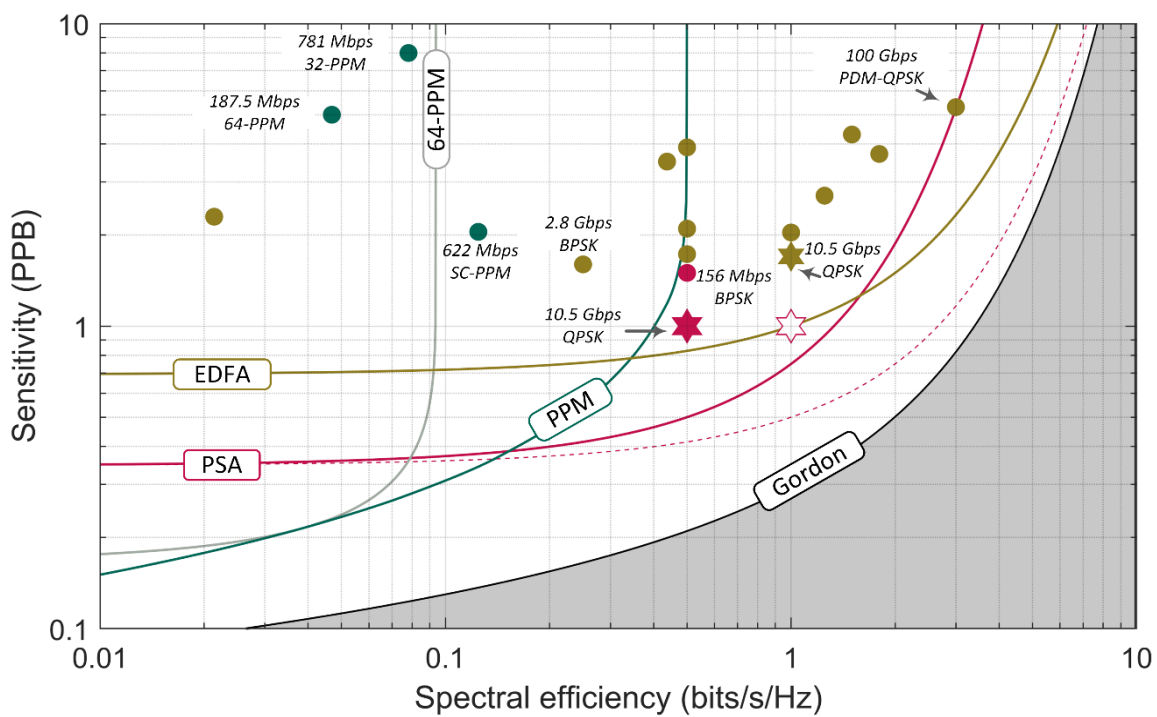

*Fig. 3. Sensitivity (photons per information bit, PPB) versus spectral efficiency (bits/s/Hz) with different implementations. Theoretical curves are indicated by lines, while experimental data are indicated with symbols. Gordon's capacity limits for reliable transmission of information including quantum effects (black) (*19*), the shaded gray area is thus fundamentally inaccessible; Capacity of DQ-coherent homodyne detector with PSA pre-*

*amplifier (red), PSA with no loss in spectral efficiency due to idler (dashed red) and EDFA pre-amplifier (brown); Envelope of PPM capacities (3) (green) and capacity of 64-PPM (grey). Experimental sensitivity records of photon counting receivers (measured in incident PPB, i.e. "black-box" sensitivity) with PPM technology for net data rates > 100 Mb/s (green markers); Record sensitivities of advanced modulation formats with pre-amplified coherent receivers for net data rates > 100Mb/s (brown markers), single quadrature detector (red marker); Experimental data was extracted from the references (8–13, 20–22); The PSA result presented here is denoted by a red star (Red filled, and unfilled) and the EDFA result is represented by brown star.*

Fig. 3. depicts the trade-off between spectral efficiency and sensitivity for receivers used in free-space communications along with experimental sensitivity records using these techniques. PPM is plotted as the envelope of all m-ary PPM (green line) showing the best achievable sensitivity for given spectral efficiency. A specific example format, 64-PPM is also plotted, as this is frequently used in space communications. Although PPM formats provide the best possible sensitivity at very low spectral efficiencies, they require large receiver bandwidth to achieve high information rates, which is very challenging with photon counting receivers.

Coherent receivers with PSA pre-amplifiers do not only have 3 dB sensitivity advantage over EDFA pre-amplifier-based receivers at low spectral efficiency but are also much more spectrally efficient than PPM formats. PSAs pre-amplified coherent receivers are amplifying both quadratures of the signal and reach the best sensitivity among all receivers over spectral efficiencies ranging from 0.16 b/s/Hz to 1.6 b/s/Hz. Considering state-of-the-art signal bandwidths of 60 GHz, this corresponds (in the ideal case) to data rates between 9.6 and 96 Gbit/s which is extremely relevant for future space communication systems.

The theoretical lines of PSA and EDFA crossover as EDFAs provide better sensitivity at high spectral efficiencies (>1.6 b/s/Hz) compared to PSAs as PSA require to transmit signal and idler thus twice the bandwidth. However, for single channel systems, as employed in space communications, receiver bandwidth utilization is more important than optical spectral efficiency due to the unrestricted channel bandwidth and limited receiver bandwidth. In this case the loss of spectral efficiency due to the idler can be ignored as at the same symbol rate the receiver bandwidth of PSA and EDFA pre-amplified receiver are the same. The PSA curve then shifts towards the right by 3 dB as indicated by the red dashed line in Fig. 3. The result if we also ignore the loss in optical spectral efficiency in the experiment is indicated by the white star.

It should be noted that the PSA achieves the same theoretical sensitivity as an unamplified ideal single quadrature (SQ) detector with 100% quantum efficiency and without thermal noise (*19*), which is known to provide the best sensitivity among non-photon counting receivers. However, PSAs provide noise-free amplification for both quadratures of the signal, resulting in twice the data-rate for the same receiver bandwidth. In addition, the amplification eliminates the impact of thermal noise and limited quantum efficiency making it more realistic to reach the fundamental sensitivity limit experimentally.

Our experimental result (1 PPB) is approximately 3 dB above the theoretical sensitivity at the spectral efficiency (0.5 b/s/Hz). The factors causing this penalty include the noise figure of PSA (1.2 dB), implementation penalty of QPSK transmitter and receiver (0.4 dB), DVBS2 limit (0.7 dB from Shannon limit), losses of WDM couplers (0.2 dB), presence of pump power and penalty due to phase noise added by the injection locking mechanism (0.3 dB).

We conclude by noting that a "black-box" record sensitivity of 1 PPB was demonstrated at 10.5 Gbps using a simple, spectrally efficient modulation format, enabled by the PSA and by ultra-low power injection locking based pump recovery. This sensitivity is achieved by approaching the fundamental limit of coherent reception of optical signals with a novel noise-free phase sensitive pre-amplifier. PSA pre-amplified coherent homodyne receivers are the most sensitive receivers in the practical range of spectral efficiencies between 0.16 and 1.6 bits/s/Hz and in comparison, to solutions relying on spectrally inefficient PPM formats, result in an order of magnitude better receiver bandwidth utilization. The fundamental advantages enable reach extension, increase of information rate and/or reduction of size of the involved optics and we believe that these results represent a significant contribution in the field of space communication and LIDAR applications such as Earth monitoring.

**Acknowledgements**

We thank A. Mecozzi, C. Antonelli, A. Lorences-Riesgo, K. Kikuchi, M. Mazur and M. Karlsson for discussions. Sumitomo Electric and OFS Denmark are acknowledged for providing optical fibers used in the experiments.

**Funding**

This work was supported by the Swedish Research Council (grantVR-2015-00535) and European Research Council (project ERC-2018-PoC 813236).

**Author contributions**

R.K performed the experiment with inputs from J.S and P.A.A, J.S worked on DSP. R.K worked on FEC and wrote the manuscript together with P.A.A and J.S.

**Competing interest**

P.A.A. has filed a patent application together with co-inventor Samuel L.I. Olsson, application number US201715540289 (pending). The conceptual content in the application is similar to the description in this paper, but it does not contain any experimental or theoretical results

**Data and Material availability**

All data is available online in a public repository, can be accessed using the link
https://data.mendeley.com/datasets/76vxmfgz3d/7